\documentstyle[11pt]{article}
\catcode`\@=11

\input{epsf}
\ifx\epsffile\undefined
\newlength{\epsfysize}
\def\epsffile#1#2#3#4]#5{}
\fi

\begin{document}
\begin{titlepage}
\samepage{
\setcounter{page}{1}

\font\fortssbx=cmssbx10 scaled \magstep2
\hbox to \hsize{
\hbox{\fortssbx University of Wisconsin - Madison}
\hfill$\vcenter{\normalsize\hbox{MADPH-98-1042}
	        \hbox{UCD-97-27}
	        \hbox{hep-ph/9802422}
                \hbox{February 1998}}$ }

\vspace{0.8in}

\begin{center}

{\Large{\bf Direct Messenger-Matter Interactions in\\
Gauge-Mediated Supersymmetry Breaking Models}}

\vfil
{\large{Tao Han and Ren-Jie Zhang\\}}
\vspace{.25in}
 {\it  Department of Physics\\
       University of California\\
       Davis, CA 95616 USA\\
       and\\
       Department of Physics\\
       1150 University Avenue\\ 
       University of Wisconsin\\
       Madison, WI 53706 USA\footnote{Present address.}}
\end{center}
\vfil

\begin{abstract}
{\rm We categorize general messenger and matter interactions 
in gauge-mediated SUSY breaking models 
by an $R$-parity for the messengers
and study their phenomenological consequences.
The new interactions may induce baryon- and lepton-number
violating processes as well as flavor-changing 
neutral currents. Bounds on the couplings from 
low-energy data are generally weak due to
the large messenger mass suppression, except for
the constraint from proton decay.
The soft masses for the scalar particles receive
negative corrections from the new interactions.
Consequently, in certain region of SUSY parameter space
the $\mu$-parameter is greatly reduced.
The pattern of radiative electroweak symmetry breaking,
SUSY particle mass spectrum and decay channels 
are also affected, leading to observable experimental
signature at the current and future colliders.}
\end{abstract}
\vfil }
\end{titlepage}

\newcommand{\gsim}{\lower.7ex\hbox{$\;\stackrel{\textstyle>}{\sim}\;$}}
\newcommand{\lsim}{\lower.7ex\hbox{$\;\stackrel{\textstyle<}{\sim}\;$}}
\newcommand{\GeV}{{\rm GeV}}
\newcommand{\TeV}{{\rm TeV}}
\newcommand{\La}{{\Lambda}}

\section{Introduction}

A model with gauge-mediated supersymmetry breaking (GMSB) \cite{DN} 
is a simple and well-motivated version of the minimal supersymmetric
extension of the Standard Model (MSSM). In addition to the
observable sector and a supersymmetry (SUSY) breaking hidden 
sector, the model also possesses messenger fields which 
mediate the SUSY breaking to 
the observable fields via the SM gauge interactions. 
The ``minimal'' model has a pair of messengers
transforming under the SU(5) representation ${\bf 5+{\bar 5}}$,
decomposed as color triplets ($D+\overline D$) and weak doublets
($L+\overline L$). They couple to a gauge singlet field $S$
through a superpotential
\begin{equation}
W_{\rm minimal} = \lambda ( S D \overline D + 
S L \overline L ),
\label{mini}
\end{equation}
where $S$ acquires non-zero vacuum expectation values for both
its scalar component ($\langle S\rangle$) and auxiliary component ($F$).

There are several attractive features in this minimal model.
First, all supersymmetric particle masses are determined 
by two parameters: the messenger scale $M=\lambda\langle S\rangle$ 
(the messenger fermion mass) and the effective
SUSY breaking scale $\Lambda=F/\langle S\rangle$. 
The gaugino and scalar soft 
masses are given, at one- and two-loop level respectively, 
by \cite{DN}
\begin{eqnarray}
  M_i(M) &\simeq&  {\alpha_i(M)\over 4\pi}\La \ ,
      \qquad i=1,2,3\\
  {\tilde m}^2(M) &\simeq& 2 \sum_{i=1}^3 C_i 
  \left({{\alpha_i}(M)\over 4\pi}\right)^2 \La^2 \ ,
\label{gm}
\end{eqnarray}
where $C_i$'s are $4/3$, $3/4$ 
for the fundamental representations of SU(3), SU(2) and $3Y^2/5$
for ${\rm U}_{\rm Y}(1)$. If $\La \sim {\cal O}(10-100$ TeV),
the SUSY particles (sparticles) can have a desirable mass 
spectrum of ${\cal O}$(100 GeV). Second, since
the scalar masses are degenerate in the family
space, the flavor-changing neutral current (FCNC) 
and CP-violation in SUSY sector are  
generally small. Finally, the gravitino mass is typically 
of $\cal O$(keV) (assuming only one SUSY breaking scale),
so it is the lightest supersymmetric particle 
in this model. Consequently, the lightest neutralino
promptly decays to a gravitino plus a photon via an enhanced 
gravitational interaction. This would have significant
implications for collider phenomenology and particle 
cosmology \cite{PR}.

However, there may be cosmological problems in the messenger
sector \cite{DM}.
Due to the conserved ``messenger number'' in Eq.~(\ref{mini}),
the lightest messenger particle (LMP) is stable. Although
naturally neutral \cite{HH} in most of the SUSY parameter
space, the LMP would have to be lighter than a few TeV in order 
not to overclose the Universe \cite{DM,HH} in the standard
inflationary cosmology. On the other hand, 
direct searches have already excluded a scalar dark matter 
particle with a mass less than about 
3 TeV at a $90\%$ confidence level, assuming it accounts 
for more than about $30\%$ of a galactic halo with local 
density $0.3$ GeV/cm$^3$ \cite{DMdetec}.
One would have to introduce the messenger-Higgs
mixing, along with a gauge singlet, to evade the direct
dark matter detection \cite{HH}. However, with those additional
interactions, one may run into
the $\mu$-$B_\mu$ problem \cite{DN,DGP,DNS}.
Besides, such a low-mass LMP needs a certain degree 
of fine-tuning.

One possible solution to the problem
is to abandon the messenger number conservation 
by introducing direct messenger-matter interactions. 
In fact, it is natural to consider this possibility 
since the messengers intrinsically carry the SM gauge quantum
numbers. In this paper, we study general interactions
between the messengers and MSSM fields. 
In Sec.~2 we present
the most general superpotential in the framework
of the minimal GMSB model and categorize different
terms with an $R$-parity for the messengers. 
We then derive the effective Lagrangian at low 
energies and examine the current experimental 
constraints on the couplings in Sec.~3. We also 
study the theoretical implications on the SUSY 
particle mass spectrum, the electroweak symmetry breaking 
(EWSB) and the $\mu$-parameter. In Sec.~4 we make some general
remarks and draw our conclusion.

\section{General Messenger-Matter Interactions}

The most general superpotential with direct messenger-matter
interactions allowed by the SM gauge symmetry is
\begin{eqnarray}
W_{\rm mix} &=&
 H_D L_4 {\overline E}\ +\ 
 H_D Q {\overline D_4}\ +\ 
 L L_4 {\overline E}\ +\ 
 Q L_4 {\overline D}\ +\ 
 Q {\overline L_4} {\overline U}
\nonumber\\
 &+& {\overline E}\ {\overline U} D_4 \ 
  +\  L Q {\overline D_4}\ 
  +\  Q Q D_4 \ 
  +\  {\overline U}\ {\overline D}\ {\overline D_4}
\nonumber\\ 
 &+& H_U L_4 \ +\ H_D {\overline L_4}\ 
  +\ L {\overline L_4}\ +\ {\overline D}  D_4\ 
  +\ Q L_4{\overline D_4}\ ,
\label{gen}
\end{eqnarray}
where we have suppressed the Yukawa coupling constant
for each term
and the generation indices for the superfields.
A subscript ``4'' has been introduced
for the messenger fields in Eq.~(\ref{mini}) as 
$(D_4,{\overline L_4})$ and $({\overline D_4}, L_4)$,
in analogue to the three-generation matter fields.

At this superpotential level, 
the bilinear terms can be rotated away at no cost
by properly redefining the superfields, 
so we will not consider them any further. 
We will also ignore the state mixing among
the messengers and MSSM fields associated with
those rotations.
The last term in Eq.~(\ref{gen}) is
the only one which involves two messenger fields.
When examining its
physical consequences at low energies by integrating
out the heavy messenger fields, the resulting 
operators would be more suppressed. Although this 
term respects the SM gauge symmetry, it does not 
naturally arise in an SU(5) unification theory.  
We will not discuss this term further.

To classify the remaining terms in Eq.~(4), we recall
that the superfields $(H_U,H_D)$ have a positive
(matter) $R$-parity assignment while the others
$(L,{\overline E},Q,{\overline U},{\overline D})$ 
are negative. It follows that
the first two terms in Eq.~(\ref{gen}) have
different $R$-parity property from the rest.
It is therefore convenient to categorize these
two groups by their $R$-parities.
If we formally require the $R$-parity conservation,
we then can generalize the ordinary $R$-parity to 
the messengers. The two possibilities are:
messenger superfields with a positive $R$-parity, which 
we will call the $M^+$-model, and messenger superfields 
with a negative $R$-parity, the $M^-$-model.

\subsection{The $M^+$-model}

If we assign the messenger superfields 
$(D_4,{\overline L_4})$ and $({\overline D_4},L_4)$ 
with a positive $R$-parity, then $R$-parity invariance
leads to the following interaction terms
\begin{eqnarray}
W_{\rm mix}^+ &=&
y_{ij} {\overline E_i} L_j L_4 \  +\ 
y'_{ij} {\overline D_i} Q_j L_4 \ +\ 
y''_{ij} {\overline U_i} Q_j {\overline L_4}
\nonumber\\
&+&
\lambda^l_{ij} {\overline E_i} {\overline U_j} D_4 \ 
+\ \lambda'_{ij} L_i Q_j {\overline D_4}\ 
+\ {1\over 2}\lambda_{ij}^q Q_i Q_j D_4 \ 
+\ \lambda''_{ij} {\overline U_i} {\overline D_j} {\overline D_4}\ ,
\label{case1}
\end{eqnarray}  
where $y_{ij},y'_{ij},y''_{ij}$ and 
$\lambda^l_{ij},\lambda'_{ij},\lambda^q_{ij},\lambda''_{ij}$
are Yukawa couplings naturally of order one\footnote{In the 
direct-transmission model \cite{model}, the messengers
are also generally charged under the dynamical group, the 
interaction can arise from higher dimensional operators
involving the dynamical sector fields. As a result, the couplings
would generically be much smaller.},
with $i,j=1,2,3$ as generation indices. Note that
$\lambda_{ij}^q=-\lambda_{ji}^q$. According to the conventional 
$R$-parity assignment for component fields, 
$(-1)^{2S+3(B-L)}$, where $S$ is the particle
spin, $B$ and $L$ the baryon- and lepton-number 
respectively, the $R$-parity so assigned 
corresponds to that assuming 
zero $B$- and $L$-numbers for the messenger superfields.
The first three terms conserve $B$ and $L$, but generate
FCNC processes in general.
Although Eq.~(\ref{case1}) preserves $R$-parity by definition,
the $\lambda^l,\lambda'$ terms in (\ref{case1}) violate $B$ and 
$L$ by $\Delta B=1/3$ and $\Delta L=1$, and the 
$\lambda^q,\lambda''$ terms violate $B$ by $\Delta B=2/3$.
Simultaneous existence of $\lambda^l$ and $\lambda^q$ 
or $\lambda'$ and $\lambda''$ may induce abrupt proton 
decay.\footnote{One could instead assign the messengers
with appropriate $B$- or $L$-numbers to eliminate
terms inducing proton decay, regardless of the 
$R$-parity assignment. 
We thank X. Tata for a discussion on this point.}
We will discuss this point in the next section. Note that
the terms of $y_{ij}, \lambda_{ij}',\lambda_{ij}''$ are the
direct analogue to those $R$-parity violating interactions
of  $\lambda_{ijk}, \lambda_{ijk}',\lambda_{ijk}''$ in 
the MSSM.

\subsection{The $M^-$-model}

If we instead assign the messenger superfields with a negative
parity, then $R$-parity invariance excludes the terms 
in Eq.~(\ref{case1}) and we are left with the two
terms involving Higgs superfields 
\begin{equation} 
W_{\rm mix}^- =  \
   y_{i}  H_D L_4 {\overline E_i}\ 
+\ y'_{i} H_D Q_i {\overline D_4}\ .
\label{case2}
\end{equation}
This $R$-parity assignment is equivalent 
to assuming the messenger superfields $({\overline D_4}, L_4)$
to carry the same $B$- and $L$-numbers 
as the MSSM superfields $({\overline D}, L)$.
Although these two terms do not induce any $B$- or 
$L$-violating processes, the messenger couplings to
the MSSM fields will mediate FCNC processes \cite{DNS,russian}
if more than one $y_i$ (or $y'_i$) coexists.

It is interesting to note that in either $R$-parity assignment
($M^+$ or $M^-$), $R$-parity invariance forbids
the $Q L_4 {\overline D_4}$ term in Eq.~(\ref{gen}).

\section{Physical implications}

\subsection{Effective Lagrangian and Low-Energy Constraints} 

By integrating out heavy messengers, we can obtain low-energy 
effective Lagrangian in terms of $1/M^2$ expansion. 
For simplicity, we only examine the leading operators 
by assuming $\Lambda/M\ll 1$ and the sparticles
and Higgs bosons to be much heavier than the energy
scale considered. 

In the $M^+$-model of Eq.~(\ref{case1}), the first three terms 
($y$-couplings) result in the following four-fermion operators
\begin{eqnarray}
{\cal L}^+_y &=& {y_{ij}y^*_{i'j'}\over 2 M^2} 
\biggl(\overline {e^{}_{iR}}\gamma^\mu e^{}_{i'R}\ 
\overline {e^{}_{j'L}}\gamma_\mu e^{}_{jL}
+ \overline {e^{}_{iR}}\gamma^\mu e^{}_{i'R}\ 
\overline {\nu^{}_{j'L}}\gamma_\mu \nu^{}_{jL}\biggr)\nonumber\\
&+& {y'_{ij}y'^*_{i'j'}\over 2 M^2} 
\biggl(\overline {d_{iR}^\alpha}\gamma^\mu d^\beta_{i'R}\ 
\overline {d_{j'L}^\beta}\gamma_\mu d^\alpha_{jL}
+ \overline {d_{iR}^\alpha}\gamma^\mu d^\beta_{i'R}\ 
\overline {u_{j'L}^\beta}\gamma_\mu u^\alpha_{jL}\biggr)\nonumber\\
&+& \Biggl[{y'_{ij}y^*_{i'j'}\over 2 M^2} 
\biggl(\overline {d^{}_{iR}}\gamma^\mu e^{}_{i'R}\ 
\overline {e^{}_{j'L}}\gamma_\mu d^{}_{jL}
+ \overline {d_{iR}}\gamma^\mu e^{}_{i'R}\ 
\overline {\nu^{}_{j'L}}\gamma_\mu u_{jL}\biggr) + {\rm h.c.}\Biggr]
\nonumber\\
&+& {y''_{ij}y''^*_{i'j'}\over 2 M^2} 
\biggl(\overline {u_{iR}^\alpha}\gamma^\mu u^\beta_{i'R}\ 
\overline {u_{j'L}^\beta}\gamma_\mu u^\alpha_{jL}
+ \overline {u_{iR}^\alpha}\gamma^\mu u^\beta_{i'R}\ 
\overline {d_{j'L}^\beta}\gamma_\mu d^\alpha_{jL}\biggr) ,
\label{LM1}
\end{eqnarray}
where $\alpha, \beta$ are color indices. Similarly, 
we can obtain the effective four-fermion operators
for the last four terms ($\lambda$-couplings) 
in Eq.~(\ref{case1}) to leading order of $1/M^2$,
\begin{eqnarray}
{\cal L}^+_\lambda &=& {\lambda^l_{ij}\lambda^{l*}_{i'j'}\over 2 M^2} 
\overline {e^{}_{iR}}\gamma^\mu e^{}_{i'R}\ 
\overline {u^{}_{jR}}\gamma_\mu u^{}_{j'R}
- \biggl({\lambda'_{ij}\lambda'^*_{i'j'}\over 2 M^2}
 \overline {e^{}_{i'L}}\gamma^\mu \nu^{}_{iL}\ 
 \overline {u^{}_{j'L}}\gamma_\mu d^{}_{jL}
+{\rm h.c.}\biggr)\nonumber\\
&+& {\lambda'_{ij}\lambda'^*_{i'j'}\over 2 M^2} 
\biggl(\overline {e^{}_{i'L}}\gamma^\mu e^{}_{iL}\ 
 \overline {u^{}_{j'L}}\gamma_\mu u^{}_{jL}
+\overline {\nu^{}_{i'L}}\gamma^\mu \nu^{}_{iL}\ 
 \overline {d^{}_{j'L}}\gamma_\mu d^{}_{jL}\biggr)\nonumber\\
&+& {\lambda^q_{ij}\lambda^{q*}_{i'j'}\over 2 M^2}
\biggl(
 \overline {u^\alpha_{i'L}}\gamma_\mu u^\alpha_{iL}
 \overline {d^\beta_{j'L}}\gamma_\mu d^\beta_{jL} - 
 \overline {u^\alpha_{i'L}}\gamma_\mu u^\beta_{iL}
 \overline {d^\beta_{j'L}}\gamma_\mu d^\alpha_{jL}\biggr)\nonumber\\ 
&+& {\lambda''_{ij}\lambda''^*_{i'j'}\over 2 M^2} 
\biggl(
 \overline {u^\alpha_{iR}}\gamma_\mu u^\alpha_{i'R}\ 
 \overline {d^\beta_{jR}}\gamma_\mu d^\beta_{j'R} - 
 \overline {u^\alpha_{iR}}\gamma_\mu u^\beta_{i'R}\ 
 \overline {d^\beta_{jR}}\gamma_\mu d^\alpha_{j'R}\biggr)\nonumber\\
&+&  \epsilon^{}_{\alpha\beta\gamma}
\biggl({\lambda^l_{ij}\lambda^{q*}_{i'j'}\over 2 M^2} 
\overline {e^{}_{iR}}\gamma^\mu u^{c\alpha}_{i'}\ 
\overline {u^\gamma_{jR}}\gamma_\mu d^{c\beta}_{j'}
+ {\lambda'_{ij}\lambda''^*_{i'j'}\over 2 M^2} 
\overline {e^c_{i}}\gamma^\mu u^\alpha_{i'R}\ 
\overline {d^{c\beta}_{j'}}\gamma_\mu u^\gamma_{jL}\nonumber\\
&-& {\lambda'_{ij}\lambda''^*_{i'j'}\over 2 M^2}
\overline {\nu^c_{i}}\gamma^\mu u^\alpha_{i'R}\ 
\overline {d^{c\beta}_{j'}}\gamma_\mu d^\gamma_{jL} 
+ {\rm h.c.}\biggr) .
\label{LM2}
\end{eqnarray}

These effective operators are similar to those obtained 
from the $R$-parity violating interactions in the MSSM \cite{Rp}
and may lead to very rich physics.
The coupling coefficients ($y'$s and $\lambda'$s) 
are considered to be naturally of order one. 
However, if we consider only one term at a time
with given generation indices
in Eq.~(\ref{case1}), there is essentially
no significant experimental constraint on them because of 
the suppression by the large messenger mass 
$M\approx {\cal O}$(100 TeV). For the same reason, none of
the terms would lead to observable signature in the
current and near-future experiments.
Even when several terms with 
different generation indices coexist, 
most effects of those operators in 
Eqs.~(\ref{LM1}) and (\ref{LM2})
are still rather weak in general. For example,
the modifications on charged current universality
and on various $\tau$, $D$ and $B$ decays 
are too small to be observable unless the couplings 
$y_{ij},\lambda^l_{ij},\lambda'_{ij}>{\cal O}(100)$. 
On the other hand, there are processes from rare
and SM forbidden decays and from neutral meson mixing that can 
be sensitive to test certain operators 
with the accuracy of current and near-future experiments. 
For instance, considering $\mu \to e \gamma$ 
with one-loop diagrams via the virtual messenger
exchange, we obtain
\begin{equation}
 |y_{1j} y_{2j}|\  < 3\left({M\over 100\ \TeV}\right)^2\ .
\end{equation}
The most stringent bound on $y'$ comes from 
the $K_L$-$K_S$ mass difference:
\begin{equation}
  |y'_{12}y'_{21}|\ < 5\times 10^{-4}\left({M\over 100\ \TeV}\right)^2 ,
\end{equation} 
which is at a very interesting level to constrain the theory.
The $D^0$ mass difference constrains the operators $y''_{12}y''_{21}$, 
while the $B^0$ mass difference bounds the operators 
with the third generation index $y'_{13}y'_{31}$.
It is also interesting to note that the operators $y'_{ij} y^*_{i'j'}$
contributing to $K_L\rightarrow \ell^+ \ell'^-$ are of the nature
of a pseudoscalar current, so there is no explicit lepton 
mass dependence in the decay width, unlike the $R$-parity violating 
interactions \cite{RpK} where there is essentially no 
sensitivity to new physics for the decay $K_L\rightarrow e^+ e^-$. 
This feature may serve as a criterion to distinguish 
different models if a signal beyond the SM is observed.

The last three operators in Eq.~(\ref{LM2}) mediate proton 
decay such as $p\rightarrow e^+\pi^0(K^0), \mu^+\pi^0(K^0)$ 
and $\nu\pi^+(K^+)$. Requiring the proton lifetime to be
larger than $10^{32}\ $years puts very 
stringent bounds to them: products of two appropriate 
couplings are restricted at order of $10^{-21}$ 
for a $100$ TeV messenger. It is therefore unlikely
for the operators $\lambda^l, \lambda^q$ 
or $\lambda', \lambda''$ to coexist. 
In Table \ref{limits}, we summarize the meaningful 
bounds on the products of two different couplings 
in Eqs.~(\ref{LM1}) and (\ref{LM2}),
along with the corresponding experimental data \cite{PDG}.
Future high precision measurements would explore the 
operators to a more significant level.

In the $M^-$-model, by integrating out the messenger 
and Higgs fields, one may obtain fermionic bilinear terms.
However, they are not only suppressed by the heavy
masses $M$ and $m_h$, but also by chirality. They are 
generally small and we will not discuss them.

\begin{table}[thb]
\begin{center}
\begin{tabular}{|c|c|c|}
\hline\hline
 couplings & bounds  & low-energy data \\
\hline
$|y_{1j}y_{2j}|$, $|y_{i1} y_{i2}|$ & $3$ & 
$BR(\mu\rightarrow e\gamma)<4.9\times 10^{-11}$ \\
$|y_{11}y_{21}|$, $|y_{11} y_{12}|$ & $0.7$ & 
$BR(\mu\rightarrow 3e)<10^{-12}$ \\
$|y_{12}y'_{11}|, |y_{21}y'_{11}|$ & $10$ 
& $\mu$ Ti(Pb)$\rightarrow$ $e$ Ti(Pb) ($g_V^{I=1}<4\times 10^{-5}$)\\
$|y_{12} y'_{12}|$, $|y_{12} y'_{21}|$,$|y_{21} y'_{12}|$, 
$|y_{21} y'_{21}|$ & $0.05$ & 
$BR(K_L\rightarrow\mu^\pm e^\mp)<3.3\times 10^{-11}$\\ 
$|y_{11} y'_{12}|$, $|y_{11} y'_{21}|$ & $0.06$ & 
$BR(K_L\rightarrow e^+ e^-)<4.1\times 10^{-11}$\\ 
$|y_{22} y'_{12}|$, $|y_{22} y'_{21}|$ & $0.8$ & 
$BR(K_L\rightarrow \mu^+ \mu^-)<7.2\times 10^{-9}$\\ 
$|y'_{12}y'_{21}|$ & $5\times 10^{-4}$ & 
$\Delta M_K=3.491\times 10^{-12}$~MeV\\
$|y''_{12}y''_{21}|$ & $0.03$ & 
$\Delta M_D<1.38\times 10^{-10}$~MeV\\
$|y'_{13}y'_{31}|$ & $0.03$ & 
$\Delta M_B=3.12\times 10^{-10}$~MeV\\
$|{\rm Im}(y'_{12}y'^*_{21})|$ & $3\times 10^{-6}$ & 
$|\epsilon_K^{}|\approx 2.275\times 10^{-3} $\\
\hline
$|\lambda^l_{11}\lambda^l_{12}|, |\lambda^l_{11}\lambda^l_{21}|,
 |\lambda'_{11}\lambda'_{12}|, |\lambda'_{11}\lambda'_{21}|$ 
& $10$ 
& $\mu$ Ti(Pb)$\rightarrow$ $e$ Ti(Pb) ($g_V^{I=1}<4\times 10^{-5}$)\\
$|\lambda^l_{1j}\lambda^l_{2j}|,|\lambda^l_{i1}\lambda^l_{i2}|,
 |\lambda'_{1j}\lambda'_{2j}|,|\lambda'_{i1}\lambda'_{i2}|$ 
& $3$
& $BR(\mu\rightarrow e\gamma)<4.9\times 10^{-11}$ \\
$|\lambda'_{11}\lambda''_{11}|;\ 
 |\lambda^l_{11}\lambda^q_{12}|, |\lambda'_{11}\lambda''_{12}|$ 
 & $\sim 10^{-21}$ & 
$\tau(p\to e^+\pi^0;\ e^+ K^0)>10^{32}$ yr \\ 
$|\lambda'_{21}\lambda''_{11}|;\ 
 |\lambda^l_{21}\lambda^q_{12}|, |\lambda'_{21}\lambda''_{12}|$ 
 & $\sim 10^{-21}$ & 
$\tau(p\to \mu^+\pi^0;\mu^+ K^0)>10^{32}$ yr \\ 
$|\lambda'_{i1}\lambda''_{11}|;\ |\lambda'_{i1}\lambda''_{12}|$
 & $\sim 10^{-21}$ & 
$\tau(p\to \bar\nu\pi^+;\bar\nu K^+)>10^{32}$ yr \\ 
\hline\hline
\end{tabular}
\parbox{6.0in}{ 
\caption[]{\small Bounds on the couplings from low-energy 
experimental data \cite{PDG} (and $\mu$-$e$ conversion
from \cite{Bill}), in units of $(M/{100\ {\rm TeV}})^2$.} 
\label{limits} }
\end{center}
\end{table}

\subsection{Sparticle Mass Spectrum}

While the masses of MSSM fermions are protected from the 
radiative corrections either by a chiral symmetry 
or by a continuous $R$-symmetry, the scalar masses
squared receive large negative corrections from the
messenger-matter interactions in Eqs.~(\ref{case1})
and (\ref{case2}). They are given by, for the $M^+$-model,
\begin{eqnarray}
\delta m_{\tilde L_{ij}}^2 &=&  {1\over 16\pi^2} 
\sum_{k=1}^3 y_{ki} y^*_{kj}
M^2 u(\Lambda/M), \label{massb4}\\
\delta m_{\tilde E_{ij}}^2 &=& {1\over 8\pi^2} 
\sum_{k=1}^3 y_{ik} y^*_{jk}
M^2 u(\Lambda/M) .  \label{massb5}\\
\delta m_{\tilde Q_{ij}}^2 &=& 
{1\over 16\pi^2}\sum_{k=1}^3 (y'_{ki}y'^*_{kj}+y''_{ki}y''^*_{kj}) 
M^2 u(\Lambda/M), \label{massb1}\\
\delta m_{\tilde U_{ij}}^2 &=& {1\over 8\pi^2}\sum_{k=1}^3y''_{ik} y''^*_{jk} 
M^2 u(\Lambda/M),\label{massb2}\\
\delta m_{\tilde D_{ij}}^2 &=& {1\over 8\pi^2} \sum_{k=1}^3y'_{ik} y'^*_{jk} 
M^2 u(\Lambda/M), \label{massb3}
\end{eqnarray}
where we have ignored the $\lambda$-couplings in Eq.~(\ref{case1}),
and for the $M^-$-model \cite{DNS,russian},
\begin{eqnarray}
\delta m_{\tilde E_{ij}}^2 &=& 
{1\over 8\pi^2} y_iy_j^* M^2 u(\Lambda/M),\label{mass1}\\
\delta m_{\tilde Q_{ij}}^2 &=& 
{1\over 16\pi^2} y'_i y'^*_j M^2 u(\Lambda/M),\label{mass2}\\
\delta m_{H_{D}}^2 &=& 
 {1\over 16\pi^2}\sum_{i=1}^3 \biggl( |y_i|^2
+ 3 |y'_i|^2\biggr) M^2 u(\La/M) ,\label{mass3}
\end{eqnarray}
where the function
\begin{equation}
 u(x) = \ln (1-x^2) + {x\over2} \ln{1+x\over 1-x}\ .
\end{equation}
For small $x$, the function has an expansion $-x^4/6$, and it 
monotonically decreases as $x$ goes to $1$.

For $\Lambda/M\ll 1$ ($x\rightarrow 0$), the corrections
are suppressed by $(\Lambda/M)^2$, so that there is no significant
difference from the minimal model for sparticle spectrum.
On the other 
hand, for $\Lambda/M\simeq 1$ ($x\rightarrow 1$), 
the function $u$ is very negative and the corrections to 
the mass can be very substantial. 
Significant upper limits on the Yukawa couplings may be
obtained by requiring that these negative corrections
do not change the sign of the scalar mass squared.
In GMSB models, the slepton and
Higgs soft masses are generically smaller
than the squark soft masses, so the tightest 
bounds come from Eqs.~(\ref{massb4}), (\ref{massb5}),
(\ref{mass1}) and (\ref{mass3}). 
As an illustration, we choose
\begin{equation}
\Lambda=100\ {\rm TeV}, \quad \tan\beta=2
\quad {\rm and}\quad \mu > 0.
\label{PARA}
\end{equation}
By requiring the scalar mass squared to remain positive,
we find that typical upper bounds on $\sum_{i=1}^3 |y_i|^2$ 
in the $M^-$-model for several $M/\Lambda$ values are 
\begin{equation}
\begin{array}{ccccc}
 M/\Lambda=            & 1.25     & 5    & 10   & 30 \\
 \sum_{i=1}^3 |y_i|^2< & 10^{-3}  & 0.03 & 0.15 &  1.2
\label{lalimits}
\end{array}
\end{equation}
Similar constraints are also obtained for $\sum_{i=1}^3 |y'_i|^2$ 
and for the couplings in the $M^+$-model. Although the bounds 
obtained here depend on the model parameter $M/\Lambda$, they
are the only upper bounds available on individual couplings. 
They are therefore complementary to those extracted from 
the low-energy data in the previous section.

The negative corrections to the scalar masses squared 
can also induce miss-alignments of the fermion-sfermion
mass matrices and as a result, the flavor-changing 
neutral currents. A study \cite{russian} found that 
bounds on the products of two couplings from 
$\mu\rightarrow e\gamma$ and
$\mu$-$e$ conversion can be as strong as $10^{-5}$
for $\Lambda/M\approx 1$. However, the bounds obtained
there depend again sensitively on the parameter choice, 
and they are much looser for $\Lambda/M\ll 1$.

\subsection{Electroweak Symmetry Breaking and The $\mu$-Parameter}

One of the most important features in SUSY theories is
the radiative generation of the electroweak symmetry 
breaking (EWSB) \cite{ewsb}. At the scale $M_{\rm SUSY}$ 
where the EWSB is imposed, the Higgs soft mass squared
$m_{H_U}^2$ is approximately given by the solution to 
the one-loop renormalization group equation 
\begin{equation}                           
m^2_{H_U} (M_{\rm SUSY})\ \simeq\ 
m^2_{H_U}(M) - {3\lambda_t^2\over 8\pi^2} (m^2_{{\tilde Q}_3}
+ m^2_{{\tilde U}_3}) \ln\left( {M\over M_{\rm SUSY}}\right), 
\label{H2}
\end{equation}
where for simplicity $\lambda_b$ term has been neglected.
The large top-quark Yukawa coupling $\lambda_t$ can drag
$m^2_{H_U}$ at $M_{\rm SUSY}$ negative, thus triggers the EWSB.

\begin{figure}[thb]
\epsfysize=2.2in
\epsffile[-50 280 380 545]{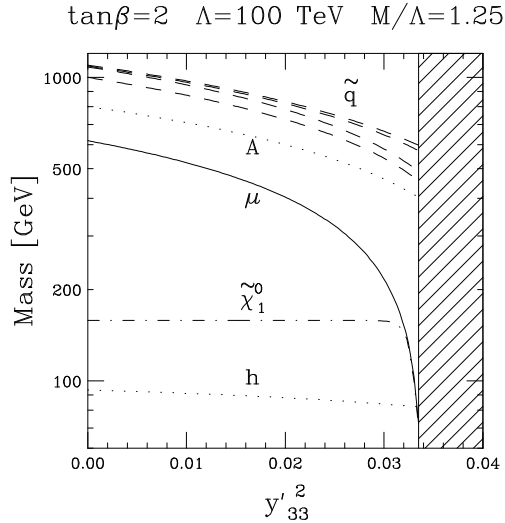}
\begin{center}
\parbox{6.0in}{
\caption[]{\small 
 Representative masses as functions of the messenger-matter
coupling $y'^2_{33}$. The hatched region does not have the right 
EWSB. Shown in the plots are the third generation squarks $\tilde q$,
the $\mu$-parameter, the lightest neutralino ${\tilde\chi}_1^0$ and
the lightest CP-even and CP-odd Higgs bosons $h$, $A$. 
\label{figure1}} }
\end{center}
\end{figure}

However, the messenger-matter interactions give
large negative corrections to the scalar masses squared,
as seen in Eqs.~(\ref{massb4})-(\ref{massb3})
and Eqs.~(\ref{mass1})-(\ref{mass3}).
When the third generation squark masses become small, 
the Higgs mass squared is less negative, and the EWSB can be 
changed significantly or may not even occur.
A more restrictive version of the EWSB condition in a 
supersymmetric theory is usually expressed in
the following (tree-level) equation which also determines the 
$\mu$-parameter
\begin{equation}
\mu^2\ =\ {m^2_{H_D}-\tan^2\beta\ m^2_{H_U}\over \tan^2\beta -1}
-{M^2_Z\over 2}\ .
\label{mu}
\end{equation}

Requiring the model to yield a desirable pattern of EWSB 
would put additional constraints on the couplings.
We consider the mass correction effects on EWSB
in $M^+$-model to Eqs.~(\ref{H2}) and (\ref{mu}).
We choose the SUSY parameters as in Eq.~(\ref{PARA}).
To maximize the effects from messenger-matter interactions,
we also choose that $M/\Lambda =1.25$ and $y'_{33}=y''_{33}$.
We have run the coupled two-loop renormalization group 
equations \cite{spectra} of soft masses to the scale 
$M_{\rm SUSY}=m_{\tilde q}^{\rm GM} + m_{\tilde q}^{\rm mix}$, where 
$m_{\tilde q}^{\rm GM}$ and $m_{\tilde q}^{\rm mix}$ represent 
the contributions to the third generation squark masses 
from the minimal GMSB model (Eq.~(\ref{gm})) and 
messenger-matter interactions (Eqs.~(\ref{massb1}-\ref{massb3})) 
respectively. At $M_{\rm SUSY}$ 
we impose the EWSB condition and calculate all physical
masses and the $\mu$-parameter consistently to the full 
one-loop order. In Fig.~\ref{figure1} we show our results
for the SUSY particle mass spectrum. The hatched region
is where the EWSB does not occur. This can be anticipated
from looking at the decreasing squark masses, which eventually
become too small to drive the Higgs mass squared of 
Eq.~(\ref{H2}) negative. The limits on the individual
Yukawa couplings obtained here are similar to that in
Eq.~(\ref{lalimits}) and are complementary to those from the 
low-energy experiments discussed in the previous section.
We note that the $\mu$-parameter decreases as the coupling
increases.

\begin{table}[thb]
\begin{center}
\begin{tabular}{|c|c|c|}
\hline\hline
({\rm GeV}) & 
$y'^2_{33} = y''^2_{33} = 0$ & $y'^2_{33} = y''^2_{33} = 0.032$ \\
\hline
$\mu$ & 619 & 152 \\
$M_1$ & 159  & 157 \\
$M_2$ & 299 & 299\\
$M_3$ & 820 & 846 \\
\hline
$m_{\tilde\chi_1^0}$ & 158  & 143  \\
$m_{\tilde\chi_2^0}$ & 316 & 169  \\
$m_{\tilde\chi_3^0}$ & 625 & 172 \\
$m_{\tilde\chi_4^0}$ & 631 & 325 \\
$m_{\tilde g}$ & 908 & 888 \\
$m_{\tilde\chi_1^+}$ & 158 & 163 \\
$m_{\tilde\chi_2^+}$ & 631 & 325 \\
$m_h$ & 94 & 83 \\
$m_A$ & 802 & 429 \\
$m_{\tilde t_1}$ & 1099 & 627\\
$m_{\tilde t_2}$ & 1000 & 495\\
$m_{\tilde b_1}$ & 1090 & 606\\
$m_{\tilde b_2}$ & 1083 & 531\\
\hline
$BR(\tilde\chi_1^0\rightarrow\gamma{\tilde G})$ & 0.90 & 0.65 \\
$BR(\tilde\chi_1^0\rightarrow h{\tilde G})$ & $\sim 10^{-5}$ & 0.26 \\
$c\tau$\ ($\mu$m) & $84$  &  $60$ \\
\hline\hline
\end{tabular}
\parbox{6.0in}{
\caption[]{\small Representative masses for the two models: 
minimal GMSB with
$y'_{33}=y''_{33}=0$, and that for $y'^2_{33}=y''^2_{33}=0.032$. 
Also shown in the Table are the $\mu$-parameter, 
the three gaugino soft masses $M_{1,2,3}$,
branching ratios for $\tilde\chi_1^0 \to \gamma\tilde G$ and 
$\tilde\chi_1^0\to h\tilde G$, and $\tilde\chi_1^0$ decay length.
}
\label{table} }
\end{center}
\end{table}

\begin{figure}[thb]
\epsfysize=2.2in
\epsffile[-50 280 380 545]{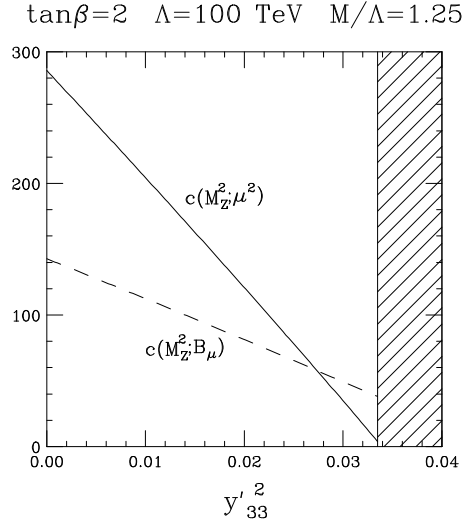}
\begin{center}
\parbox{6.0in}{
\caption[]{\small 
 Dimensionless quantities in Eq.~(\ref{finet}) which describe 
the degree of fine-tuning as functions of $y'^2_{33}$. 
\label{figure2}} }
\end{center}
\end{figure}

Models with direct messenger-matter interactions can display 
mass spectrum with very different characteristics 
from that of the minimal GMSB model. 
To demonstrate this point, we show a representative
mass spectrum in Table \ref{table}, 
where we choose $y'^2_{33}=y''^2_{33}=0.032$ 
and all other parameters are taken to be the same as 
Fig.~\ref{figure1}. For comparison, we also show that for
the minimal GMSB model with no direct messenger-matter 
interactions. The masses for the third generation 
squarks, the neutralino/chargino and especially
the Higgs bosons are all significantly lighter 
that those for the minimal model.
The lightest neutralino $\tilde\chi_1^0$ has a large 
Higgsino component, so it can decay to the light Higgs $h$ 
with a fairly large branching ratio of $26\%$. The decay length 
of $\tilde\chi_1^0$ becomes somewhat shorter as well.
These interesting features may lead to very
distinctive experimental signature in the current
and future collider experiments.
The slepton masses can be decreased significantly as 
well depending on the choice of the couplings $y'$s.

In the minimal GMSB model, because the squarks are much
heavier than other sparticles, one finds that 
$\sqrt{|m_{H_U}^2|}$ and therefore
$\mu$ are typically much larger than $M_Z^{}$.
This renders a difficult balance of Eq.~(\ref{mu})
and is usually referred to as the fine-tuning problem 
in GMSB models \cite{GMfine}.
In the presence of messenger-matter interactions, one should 
expect that $\mu$ is generally smaller (as seen from 
Fig.~\ref{figure1}), 
and the fine-tuning problem should be less severe.
We examine the dimensionless quantities 
as a measure of fine-tuning \cite{finetun}:
\begin{equation}  
c(M^2_Z;\mu^2) = |{\partial\ln{M^2_Z}/\partial\ln\mu^2}|,\ 
\quad {\rm and} \quad
c(M^2_Z; B_\mu) = |{\partial\ln{M^2_Z}/\partial\ln B_\mu}|.
\label{finet}
\end{equation}
where $B_\mu$ is the bilinear soft Higgs mass parameter.
The results are shown in Fig. \ref{figure2}. 
Indeed the fine-tuning improves for a bigger 
coupling $y'_{33}$.

\section{Discussions and Conclusion}

Before we draw our conclusions, several remarks are in order.
First, in constructing the low-energy effective Lagrangian
in the previous section, we have ignored terms proportional 
to $\La/M$. This is the case where the SUSY breaking effect
in the messenger sector is much smaller than the 
messenger scale $M$ itself. The terms proportional to
$\La/M$ have essentially the same structure as those
in Eqs.~(\ref{LM1}) and (\ref{LM2}), with somewhat
different combination of the Yukawa couplings.
The physics implications are however very much similar.

Second, in principle, one can integrate out only the heavy
messengers and obtain effective Lagrangians involving external
sparticles. For example, terms in Eq.~(\ref{case1})
could induce pair productions of $\tilde l_i \tilde l_j^*$,
and $\tilde q_i \tilde q_j^*$ at lepton and hadron colliders,
and those in Eq.~(\ref{case2}) would give Higgs and
Higgsino pair production. Although the strength
of the new interactions is generically small, these 
distinctive processes may provide new
experimental signature at future colliders.

Third, we have neglected the complication 
of the CKM matrix when deriving the couplings
from Eq.~(\ref{case1}) to Eqs.~(\ref{LM1}) and 
(\ref{LM2}). It can be systematically included
by performing the proper quark field rotation
between the weak and mass eigenstates.

Finally, although we only concentrate on a pair 
of ${\bf 5+{\overline 5}}$ in this paper,
the analysis can also be readily carried out for the cases of several 
${\bf 5+{\overline 5}}$ pairs or a pair of ${\bf 10+{\overline{10}}}$.
In the case where the messengers are a pair of 
${\bf 10+{\overline {10}}}$, 
the LMP is charged and cannot be stable without
causing problems in the standard inflationary
cosmology. The direct messenger-matter 
interactions thus may necessarily occur.
We should note that in these cases general mixing
among messengers are also possible. Under certain assumptions,
it is shown in Ref.~\cite{Dterm} that hypercharge
$D$-term contributions to the scalar particle masses 
can be generated at two-loop level,
but these terms are generally much smaller than the  
one-loop contributions from the messenger and matter interactions.

In conclusion,
we have constructed the direct messenger-matter interactions
in the minimal GMSB model. The new
interactions avoid the cosmological problem associated 
with the stable messenger particle, but they generally introduce
$B$ and $L$ violating and FCNC processes.
We obtain the low-energy effective Lagrangians by
integrating out the heavy messenger fields as well
as the sparticles. If we assume that the couplings 
are naturally of order one, we find that the constraints
from the low-energy data are generally not very restrictive
except for those leading to proton decay. On the
other hand, certain combinations of the couplings
may contribute to some flavor violating processes
within the current and future experimental reach.
We also show that the new interactions have negative
contributions to the scalar particle masses.
Consequently, one can generally reduce the value of the 
$\mu$-parameter and greatly change the pattern of EWSB.   
The fine-tuning problem associated with the $\mu$-parameter
can be alleviated at most of the parameter space.
The significantly lighter mass spectrum for the 
sparticles and Higgs bosons and the 
different sparticle decay pattern can result in distinctive
experimental signature at the current and future colliders.

{\noindent \bf Acknowledgment}

We would like to thank 
K. Cheung and R. Hempfling for helpful conversations and
K. Matchev for comments on the manuscript.
This work was supported by DOE grants 
DE-FG03-91ER40674 and DE-FG02-95ER40896.

\end{document}